# Effect of incoming radiation on the non-LTE spectrum of Xe at Te=100eV.

Marcel Klapisch[1] and Michel Busquet

ARTEP Inc., Ellicott City, MD 21042, USA.

keywords: Laboratory Astrophysics, non-LTE, radiation driven plasmas.

**Abstract**
The effect of a diluted Planckian radiation field on a Xe gas at the electron temperature of 100eV is investigated within the framework of a Collisional Radiative Model, using the HULLAC code. The atomic model spans 19 charge states, includes 20,375 configurations and contains more than $2 \times 10^6$ levels. We have simulated detailed spectra comprising more than $10^9$ transitions with the Mixed UTA model.

The radiation temperature $T_r$ is varied from 0 to 1.5 $T_e$. The dilution factor, D, applied to decrease the radiation field, is varied independently from 0 to 3 at fixed $T_r$ = $T_e$ . In both cases, the average charge state $Z^*$ increases from 15 to 27, but in different ways. It is shown that even a dilution D=0.01 changes $Z^*$ by more than 1.5. Different combinations of $T_r$ and D yielding exactly the same $Z^*$, may give line ratios sufficiently different to be observed. This fact is explained by the interplay of the shape of the radiation field and the atomic structure.

1. **Introduction**
Radiation plays an important, and sometimes fundamental, role in the state of many astrophysical plasmas. For instance, understanding the role of radiation in supernova explosions is a challenge. Further, constructing laboratory experiments that provide scaled emulation of supernova hydrodynamic phenomena is very challenging. Indeed it might seem that laboratory simulations are impossible because of the enormous difference in the scales. However, scaling arguments were used to study Rayleigh-Taylor instabilities [1] and radiative shocks [2] by comparing numerical simulations of astrophysical conditions and laboratory experiments involving light materials. The criteria for scaling, discussed in detail by Ryutov [3], indicate that heavy atoms can be used for laboratory astrophysics experiments. Following the pioneering work of Bozier [4], several experiments involving xenon were performed.[5-7] More recently, Bouquet [8], justified the use of xenon for radiative shocks studies by detailed investigations of the scaling relations.

In the present work, we study how the incoming radiation changes the charge state of xenon, and whether these changes can be detected by spectroscopic measurements on laboratory experiments. We chose plasma parameters that are not related to any particular experiment, but with typical values ensuring that without radiation the system is <u>not</u> in local thermodynamical equilibrium (non-LTE condition). Many definitions and formulas for studies of these conditions can be found in the work of

---

[*] Corresponding author; email: klapisch@artepinc.com Address: ARTEP,inc. 2922 Excelsior Springs court, Ellicott City, MD21042, USA

Rose [9]. Radiation effects in non-LTE plasmas were previously described with the FLYCHK code of Chung and Lee [10]. Hill and Rose, [11] using the code ALICE have stressed the importance of using a complete set of configurations for the atomic description used in the collisional radiative model (CRM). For the level of detail we are seeking to achieve, we used an improved version of HULLAC [12, 13]. With this version of HULLAC, we are able to explain why some lines are sensitive to incoming radiation while others are not. In section 2, we describe the recent improvements to HULLAC, and the atomic model for these very extensive computations. The results are presented in Sec. 3, followed by a discussion of the accuracy in Sec. 4. Sec. 5 provides a summary and conclusions.

2. **Computation details**.

The plasma parameters chosen for this study were electron temperature $T_e$ = 100eV and electron density $N_e$ = $10^{20}$ cm$^{-3}$. Only the radiation field, characterized by a radiation temperature $T_r$ and a dilution factor D, was varied as described in Sec. 4. The challenge – and the interest – of this case study lies in dealing with the large differences of average charge state Z* between the pure non-LTE, i.e., $T_r$ =0, and the LTE case, i.e., the case where $T_r$ = $T_e$ . The difference in Z* can be more than 10 charge states. In order to obtain reliable ion charge distributions, the computation of 19 charge states was necessary. In the majority of them, the N shell, *n=4*, is occupied by several electrons. Consequently, we encountered many configurations with numerous levels, and strong configuration interaction effects. The resulting lengthy computations were made practical by taking advantage of several modifications recently introduced in HULLAC. [12]

2.1 **Recent improvements to the code:**
(a) The subroutines for computing photo-ionization and photo-excitation rates were introduced in HULLAC for the purpose of comparing with the experiments of Bailey.[14, 15] The definitions of these rates are well known and can be found, e.g., in Refs. [9, 16]. In HULLAC, the radial integrals are computed using the efficient phase-amplitude algorithm [17].
(b) The new version of HULLAC can reproduce an incoming radiation as a combination of blackbody Planckian distributions with different radiation temperatures, each multiplied by a dilution factor:

$$I(v)dv = \sum_k D_k \frac{2hv^3}{c^2} \frac{dv}{\exp(hv/k_B Tr_k)-1} \quad (1)$$

(b) The code includes several configuration average options, where "configuration" will always refer to relativistic (jj) configuration:
(i) Relativistic configuration average, which is straightforward as HULLAC is based on wavefunctions that are solutions of the Dirac equation.
(ii) Non-relativistic configuration averages, which is obtained by the statistical-weight average of Relativistic configuration average;
(iii) Averaging detailed levels to non-relativistic configurations or superconfigurations.

(c) Mixed Unresolved Transition Array (MUTA) model [18]. In this model, the complete fine-structure, level-to-level, transition array is computed, including configuration interaction when necessary. Then any transition that is more intense than a given percentage τ of the total array is set aside as an individual line, while all the others are lumped together in an asymmetrical Gaussian distribution of intensity.

2.2 **Choice of configurations:**
As shown by other authors [11, 19], it is important to check for completeness of the atomic model. This means that the model not only includes a large number of levels, many of which are doubly-excited levels, but also that the choice of the latter is such that ionization and recombination channels exist between them and neighboring ions [20]. Here we proceed by trial and error, until adding more configurations does not change the shape of the charge distribution for each value of $T_r$. We included enough charge states so that the smallest ion fraction in the distribution in all cases was smaller than $10^{-6}$.

In the case of n=4 open-shell ground configurations, the following four combinations were considered. For singly-excited states, we included states: (1) with one excited orbital in the same complex and (2) with one excited orbital in the n=5 or n=6 shells. In addition, we included the following doubly-excited states: (3) two excited orbitals in the n=4 complex, and (4) one excited orbital in the same n=4 complex along with one on the n=5 or n=6 shells. For the highest charge states, the n=3 shell was open, and a similar pattern as used with the n=4 case was adopted. At this stage, all the computations were done in the relativistic average configuration mode, which is quick and sufficiently detailed. Once the configuration basis was chosen, we again computed every ion with the MUTA option. This consists of detailed level accounting of all levels including configuration interactions for energy levels and radiative transitions. Note that the MUTA algorithm was used to generate the radiative transition arrays, while the collisional processes and photo-ionization continued to be described on an average configuration basis.

As a result of this process, we included all the ions from $Xe^{11+}$ to $Xe^{29+}$, totaling 20,375 configurations and 2,070,392 fine structure levels. Fig. 1 shows the number of detailed levels and number of relativistic configurations in these computations, for which the former is up to 500 times larger. The chemical symbols on the curves indicate the isoelectronic sequence. The minima occur near closed shell ground states.

Fig. 2 shows the number of individual transitions actually computed, along with the lumped MUTA's, the strong lines that were kept as such in the MUTA algorithm, and the configuration-to-configuration arrays. In extreme cases, the ratio of individual lines to MUTA is five orders of magnitude. With commonly accessible computational resources it would have been extremely time consuming to calculate the collisional equivalent for these transitions, to build and to solve the CRM. This is why the upper curve of Fig. 1 is limited to the ions Nb- to Co- like, and the upper curve in Fig. 2 is limited to ions 16 to 27, i.e., $Xe^{15+}$ to $Xe^{26+}$.

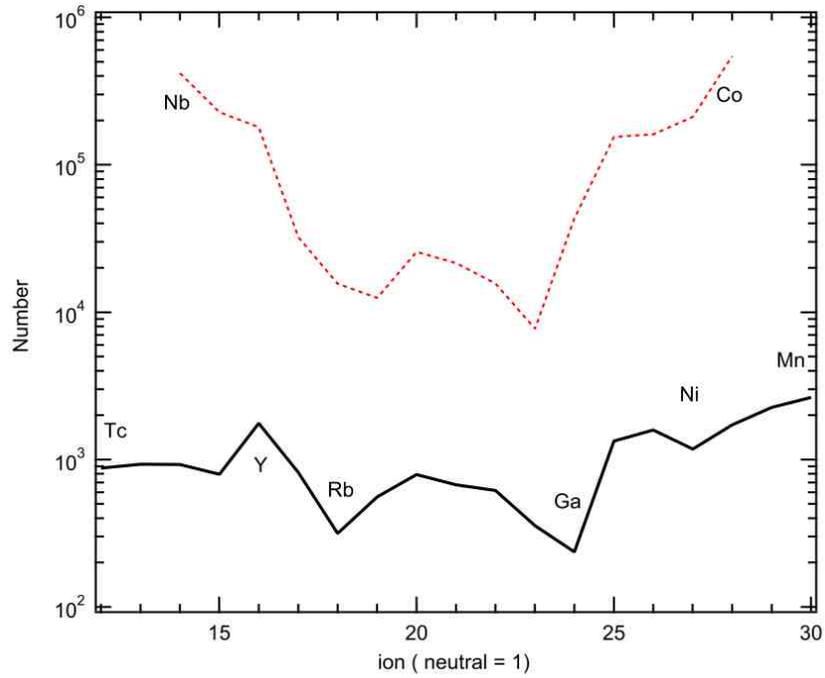

*Figure 1. Number of configurations (continuous black line) and detailed levels (dotted red line) for each Xe ion. The ion numbering starts from 1 (neutral), as in spectroscopic notation. The chemical symbols indicate the isoelectronic sequence of these ions.*

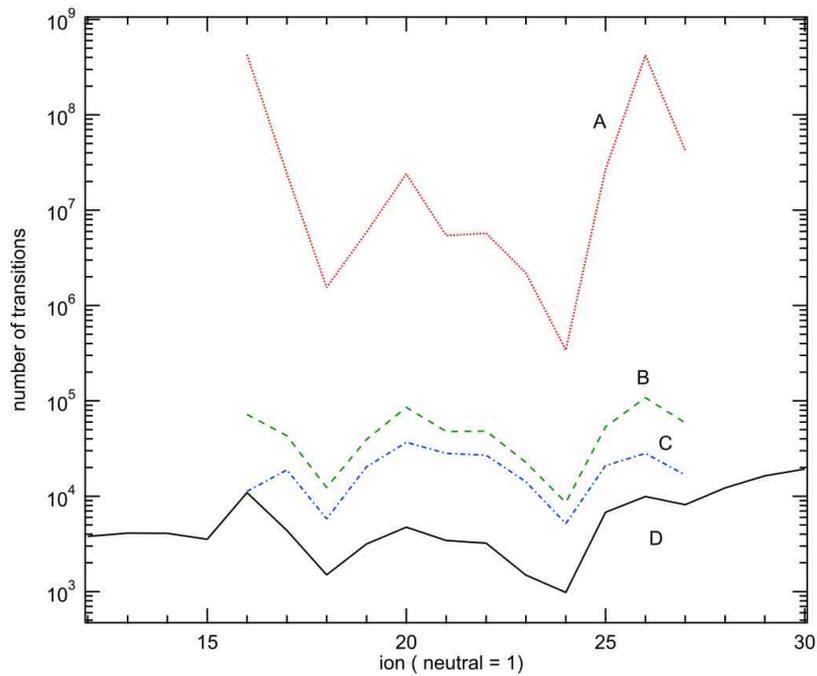

*Figure 2. Number of transitions. Ion notation same as in Fig.1 –neutral is 1. A: level-to-level transitions, B: MUTA's, C: strong lines not included in the MUTA's, D: configuration-to-configuration transitions.*

## 3. Results.

For this case study, the radiation field was described by one radiation temperature $T_r$ and one dilution factor D, which were varied independently. As mentioned in Sec. 2, the electron temperature $T_e$ and the electron density $N_e$ were fixed respectively at 100 eV and $10^{20}$ cm$^{-3}$. Once the energies and the rates for all ions are computed, our CRM solver reads the relevant files and solves the rate equations for different radiation temperatures and/or dilution factors.

### 3.1 Variation of $T_r$

Fig. 3 displays the change in the charge distribution N(ion) versus the ion charge for different radiation temperatures and a dilution factor D = 1. Note that on the figures the notation follows the spectroscopic convention, i.e., neutral = 1.   At $T_r$ =0 ( "pure" non – LTE), the distribution is broad and asymmetrical. With increasing $T_r$, the distribution gets narrower and more symmetrical. One sees that even with a $T_r$ of 10 eV, a change in the distribution becomes visible. Note that the CRM here can handle a radiation dominated case with $T_r$ = 150 eV, i.e. 1.5 x $T_e$. The only limit is the number of ions in the model. The effect of varying $T_r$ on emission spectra is shown on Fig. 4. The spectra were vertically offset for clarity. The transitions around 100 – 120 eV belong to $\Delta n=0$ in the $n=4$ complex and these transition energies are nearly Z*-independent, but are strongly affected by configuration interaction. Even though they occur at around the same energy, it is clear that the line intensity ratios are very different for the different $T_r$ . The transitions around 175 – 220 eV are $\Delta n=1$ (*4* to *5*) and those at 230 – 300 eV are $\Delta n=2$ *(n=4* to *n=6)* transitions, whose energy varies as Z$^2$.  Not only the energies of those arrays vary with $T_r$, but their intensity as well.

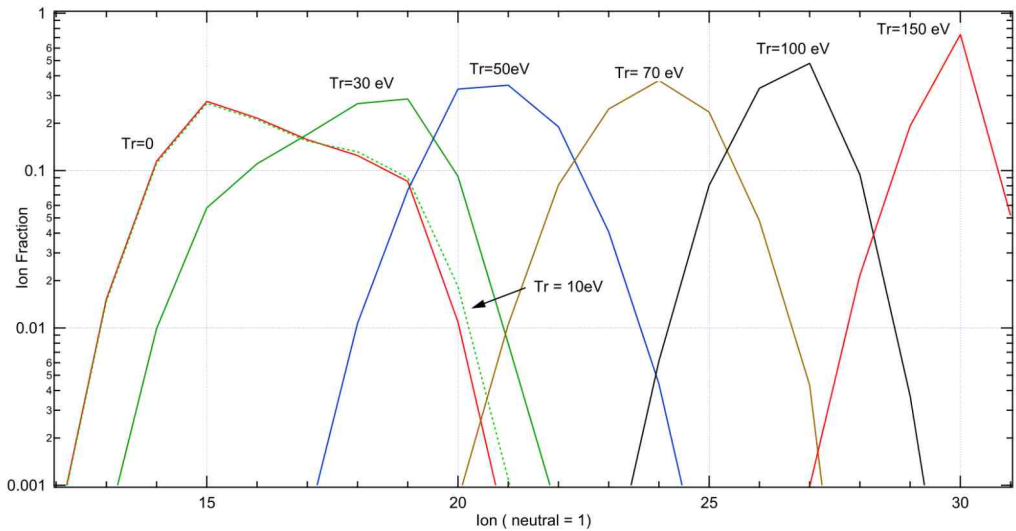

*Figure 3.  Charge distributions of Xe at $T_e$ =100eV, $N_e$ =$10^{20}$ /cm$^3$ with different $T_r$ . In all cases, the dilution factor D=1.*

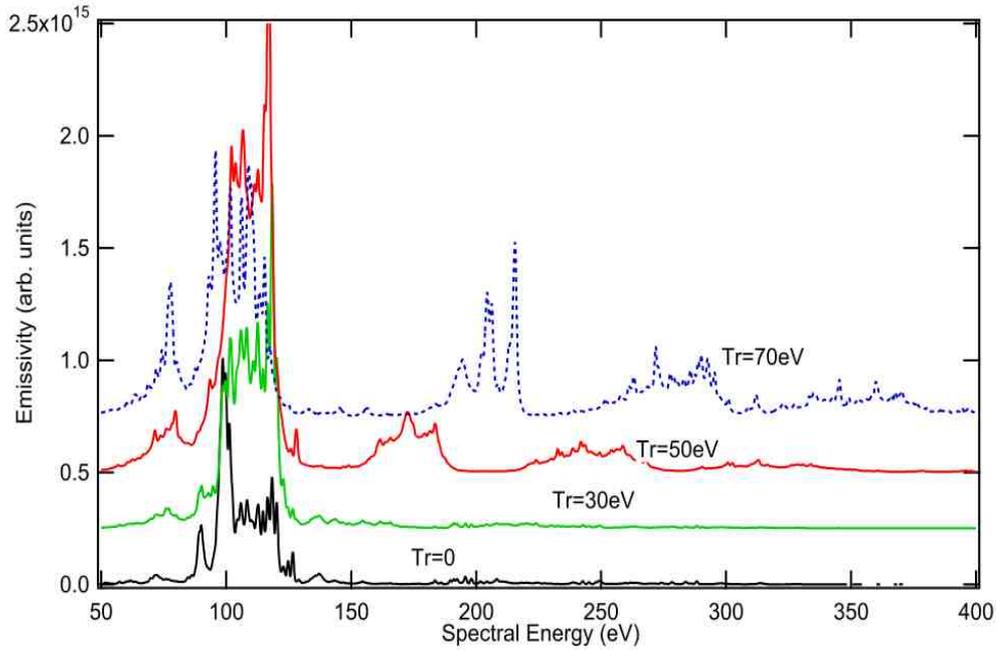

*Figure 4. Emission spectra of Xe at $T_e$ =100eV, $N_e$ =$10^{20}$ /cm$^3$ with different $T_r$. In all cases, the dilution factor D=1. The spectra were given an offset to distinguish them. The dotted line is $T_r$ =70 eV.*

### 3.2 Variation of the dilution factor

Fig. 5 shows the change in the charge distributions for different values of the dilution factor, from zero ("pure" non –LTE) to 3 – radiation dominated. In all cases the radiation temperature $T_r$ was kept at 100 eV, i.e., equal to $T_e$. Figure 5 shows that the cases D=0 and D=1 are identical to $T_r$ =0 and $T_r$ =100 of Fig. 3 respectively, as expected. For the other values of D, the distributions are wider and more asymmetrical than those in Fig. 3 and are centered at about the same average Z*. It is striking that even a dilution factor of 0.01 observably changes the ion distribution. Indeed, in Fig. 6 this difference in ion distribution translates into line intensity ratios that are different enough to be experimentally detected. These different line ratios, around 100 eV and 120 eV, are highlighted with straight lines. This means that if $T_e$ and $N_e$ could be measured independently (not by spectroscopy), then one could obtain an estimate of the radiation field. Alternatively, if one would rely on line ratios to diagnose $T_e$ and/or $N_e$ without taking into account radiation field, the result would be erroneous.

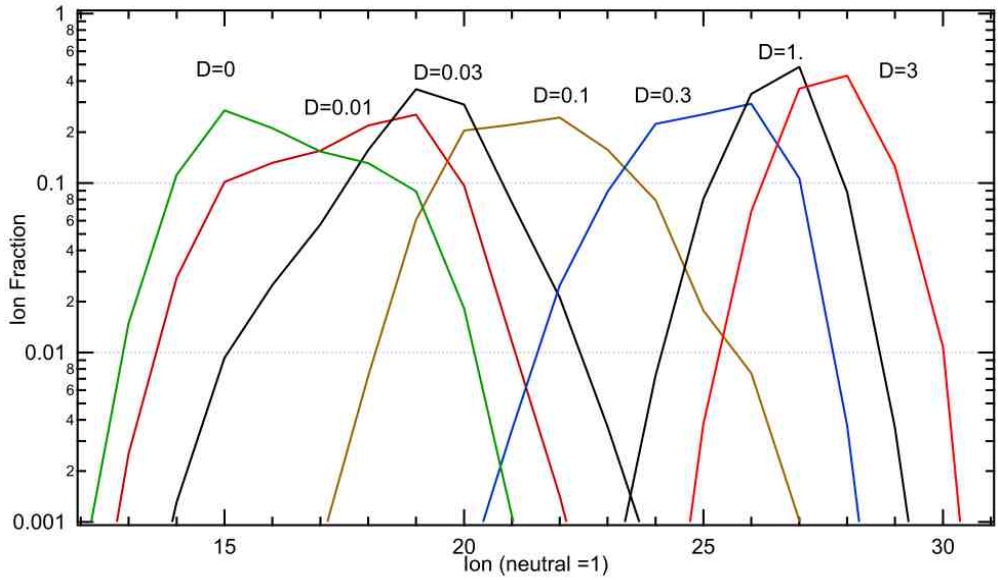

*Figure 5. Charge distributions of Xe at $T_e$ =100eV, $N_e$ =$10^{20}$ /cm$^3$ with different dilution factors D. In all cases, the radiation temperature $T_r$ = 100eV.*

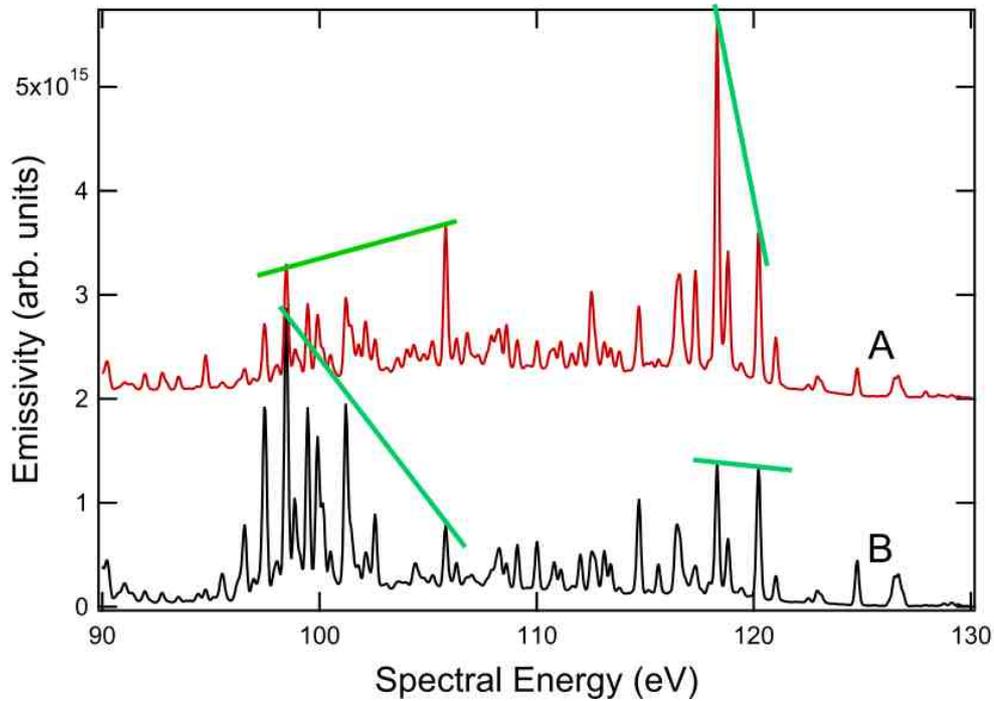

*Figure 6. Emission spectra of two cases: (A) $T_r$ =100, D= 0.01 (offset to separate the spectra) (B) $T_r$ =0.*

## 3.3 Comparing effects of $T_r$ and of D

The variation of $Z^*$ with D and with the ratio $T_r / T_e$ is displayed in Fig 7. Note that the range of variation of $Z^*$ is large and represents a motivation for carrying out this study. It shows again that $Z^*$ is very sensitive to D. $Z^*$ does not change at all for $T_r / T_e$ smaller than 0.1, and for $T_r = 20$ eV, the change in $Z^*$ is only 2.5%. The two lines cross, as expected for $T_r / T_e = 1.$, and D=1.

The shapes of the charge distributions in Figs. 3 and 5 are quite different. This is to be expected, since different shapes of the incoming radiation function *I(v)* can excite or ionize different transitions. We investigated whether this difference could be measured experimentally, for which purpose we compared two charge distributions giving exactly the same $Z^*$. The first case is $T_r = 40$ eV, D=1 (case A), and the second is $T_r = 100$eV, D=0.0324876 (case B). These two cases give the same $Z^*$ =17.98911. Fig. 8 shows the charge distribution for these two cases. These curves are nearly identical, the first case being narrower, in a manner similar to the comparisons between Figs. 3 and 5. Nevertheless, Fig. 9 again shows that some line intensity ratios around 100 eV are different enough to be detected experimentally.

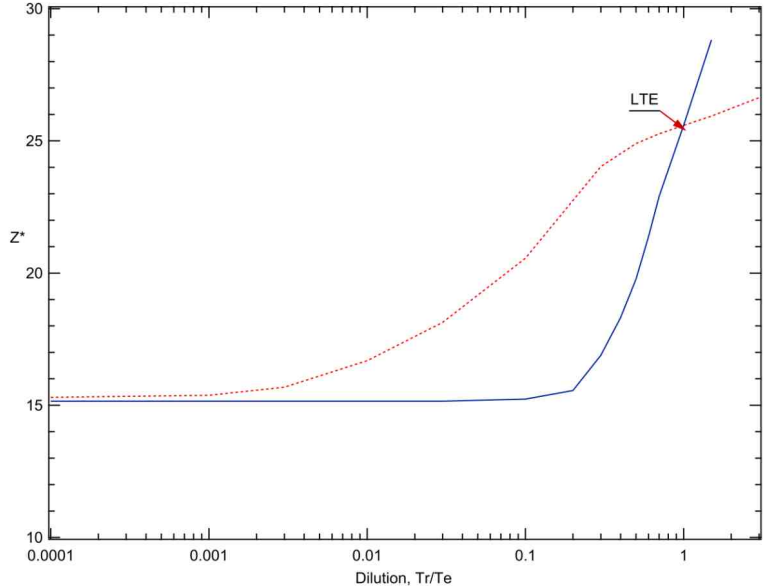

Figure 7. Evolution of $Z^*$ as a function of the dilution factor for $T_r = 100$ eV (dotted line), and as a function of the ratio $T_r / T_e$ (solid line) for D =1. The lines cross when LTE conditions prevail, i.e., $T_r / T_e = 1$, $T_r = 100$, and D=1.

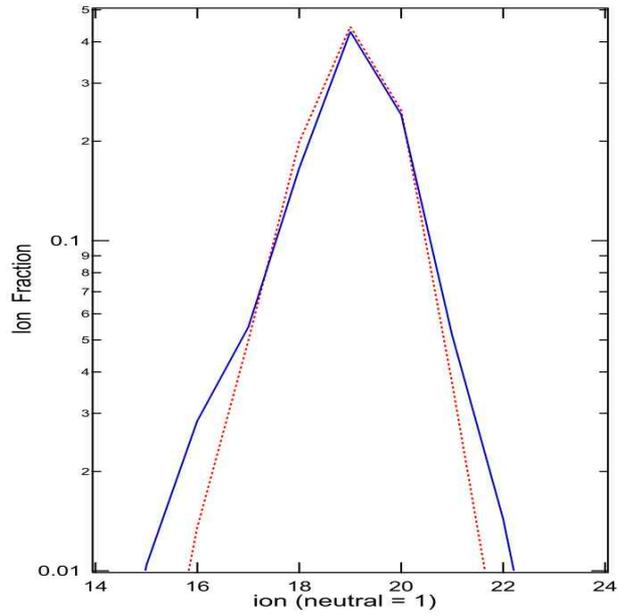

*Figure 8. Charge distributions in two cases with exactly the same Z\* = 17.89811 . (A) $T_r$ =40eV, D=1.; (B) $T_r$ =100 eV, D=0.0324876.*

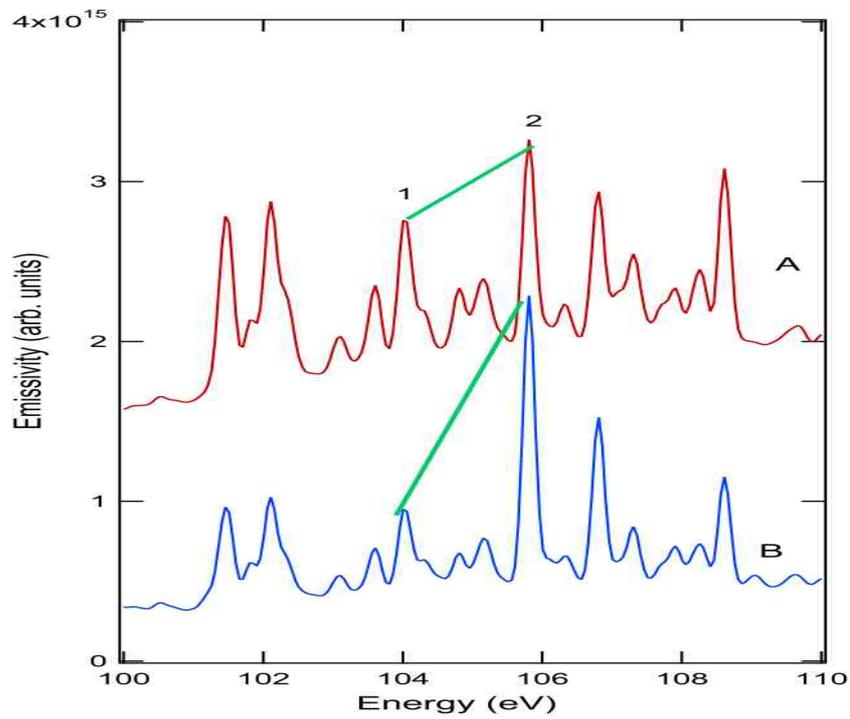

*Figure 9. Emission spectra of the two cases of figure 7. (A) $T_r$ =40eV, D=1., (B) $T_r$ =100 eV, D=0.0324876.*

## 3.4 Mechanisms of photo-excitation

An interesting example is the case of two features at 104.0 eV (line 1) and 105.8 eV (line 2) because they belong to the $Xe^{18+}$ ion . Both features are actually very narrow MUTA's connecting the levels of their respective configurations. Their intensity ratios are 1.55 in the case $T_r$ = 40 eV, D=1. (case A), and 3.48 in the case $T_r$ = 100eV, D=0.0324876 (case B). Figure 10 shows the contribution of $Xe^{18+}$ to the emission spectrum. In addition to the two cases A and B as above, the case C, with no radiation, is shown. In each case, the spectrum was normalized by the proportion of $Xe^{18+}$ in the appropriate charge distribution.

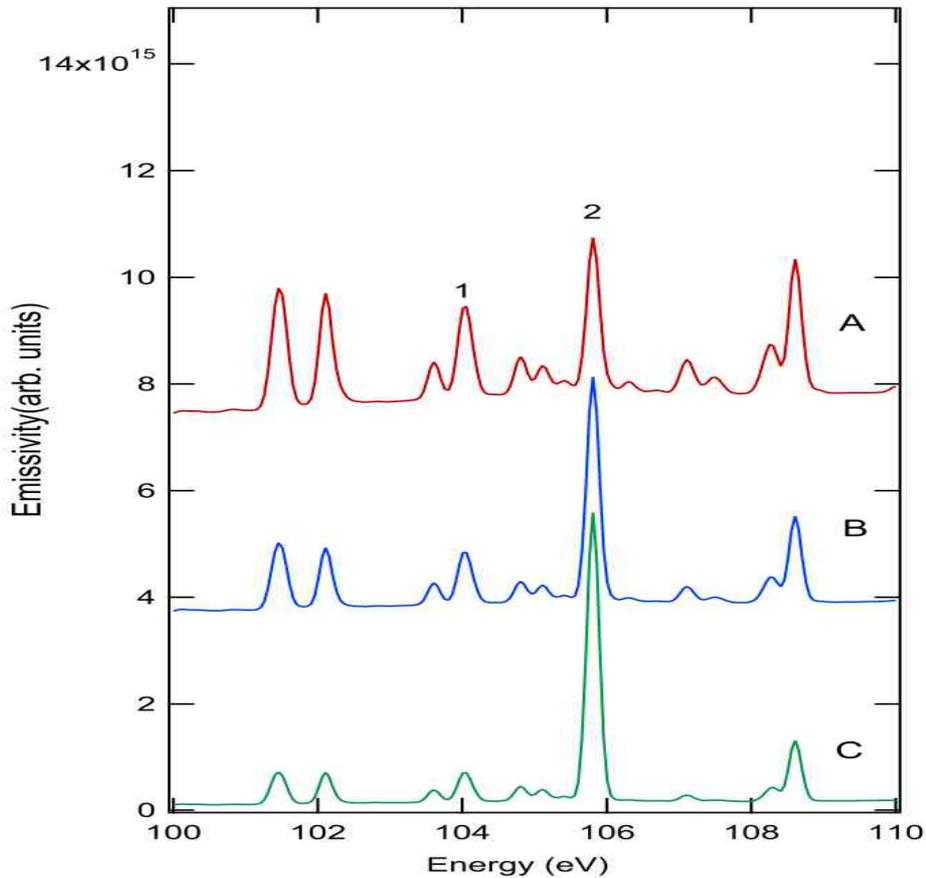

Figure 10. Contribution of $Xe^{18+}$ to emission spectra. normalized to their respective proportion in the charge distribution. (A) and (B) as above. (C) No radiation ($T_r$ =0.)

For a system with many levels subject to broadband radiation we can expect that each level will be populated in a different way. Since our model is very detailed, it was possible to analyze and explain the different behavior of these two neighboring features.

The feature at 104.0 eV is a $4s^2 4p^4 4f4d \rightarrow 4s^2 4p^4 4d^2$ transition, the upper levels of which are 278 eV above ground state. Its intensity is increased by the $T_r$ = 40 eV radiation field by nearly a factor of 2. This can be explained by the photo-excitation rate from a lower level, $4s^2 4p^5 4f$ lying at 189 eV, which is more populated than the ground state $4s^24p^6$. This pumping transition energy of 89 eV is not far from the maximum of the black body at 115 eV.

Let us define the photo-pumping ratio $R_{pp}$ as the ratio of the photo-excitation rate to the de-excitation rate for the same transition. In case A, $R_{pp}$ for $4s^24p^5f \rightarrow 4s^24p^44f4d$ is 1.75, while for case B this ratio is 0.37. In case B there is more de-excitation probability than photo-excitation because there are less photons at the pumping transition energy. Therefore, the intensity of line 1 is higher in case A than in case B.

The intensity of line 2 at 105.8eV is slightly higher in case B and C. This transition can be identified as $4s^24p^44d^2 \rightarrow 4s^24p^54d$. The upper level lies at 174 eV above the ground state. For this line, as opposed to line 1, the only configuration from which the upper level $4s^24p^44d^2$ can be photo-excited is the lower configuration of the observed transition itself, namely $4s^24p^54d$, because the neighboring configurations do not satisfy the dipole transition selection rules. The calculations show that for this line the $R_{pp}$ is 0.232 in case A and 0.058 in case B. In other words, the effect of the radiation field is to change the apparent transition probability. The slight increase in the intensity may arise from radiative cascades and from the normalization of level populations in each ion. If the population of the upper level of line 1 decreases, the upper level of line 2 may increase accordingly. The description of these two transitions are summarized in table I. Table II shows the $R_{pp}$ for the two cases A and B. In both cases, the effect of incoming radiation is larger for line 1 than for line 2 because of the atomic structure involved.

*Table I.*
*Description of configurations involved in lines (1)-104eV and line (2) 105.8 eV in figures 9 and 10.*

|  | **Line 1 at 104 eV** | **Line 2 at 105.8 eV** |
| --- | --- | --- |
| Observed transition | $4s^24p^44f4d \rightarrow 4s^24p^44d^2$ | $4s^24p^44d^2 \rightarrow 4s^24p^54d$ |
| Pumping transition | $4s^24p^5f \rightarrow 4s^24p^44f4d$ | $4s^24p^54d \rightarrow 4s^24p^44d^2$ |

*Table II.*
*Ratio of photo-excitation to de-excitation rates of line (1)-104eV and line (2) 105.8 eV in figures 9 and 10.*

| $R_{pp}$ | **Line 1 at 104 eV** | **Line 2 at 105.8 eV** |
| --- | --- | --- |
| Case A | 1.75 | 0.37 |
| Case B | 0.23 | 0.058 |

**4. Discussion**
This is a purely theoretical study of the effect of a Planckian radiation field on a non-LTE plasma at fixed electron density of $N_e = 10^{20}$ /cm$^3$, and electron temperature $T_e$ of 100 eV. It is doubtful that these conditions can be exactly reproduced in experiments. Therefore, the intention of this work, as mentioned in the introduction, is to illustrate the possible effects of a radiation field on a plasma not in local thermodynamical equilibrium.

Some inaccuracy of the charge distribution could come from the fact that the latter is computed in a model of relativistic configuration averages, thus the effect of possible metastable levels is not taken into account. Based on previous comparisons [15] we estimate the accuracy of the transition energies to be 0.1% for $\Delta n > 0$ and 1-2 % for $\Delta n = 0$. We feel that the MUTA model[18], with the level-to-level transition computed with configuration interaction (CI), should be adequate to reproduce the spectrum. Indeed, configuration mixings in the n=4 complex are very strong, with many a level having its largest component less than 50%. Thus, the shift of the mean transition energy caused by CI can be noticeable [21]. The standard UTA model [22] might lead to misinterpretation of the transitions. Finally, we note that no density effect, like lowering of the ionization potential, or optical thickness, is taken into account.

5. **Summary and conclusion**
In this work, we used the code HULLAC to explore the possible effects of a Planckian radiation field on a plasma that is in a-non-LTE state. For this purpose we chose xenon, which is widely used in laboratory simulations of scaled astrophysical phenomena[23]. A plasma with an electron temperature $T_e$ of 100 eV, and an electron density $N_e$ of $10^{20}$ cm$^{-3}$ was used. These plasma parameters, although not related to any particular experiment, are believed to be achievable. To simplify the study, we simulated the radiation field as a Planckian at different radiation temperatures $T_r$, with $0 < T_r < 1.5 \times T_e$. Calculations were also performed with a diluted Planckian, which is obtained by a dilution factor D. In the case $T_r = T_e$, D was varied with $0 < D < 3$. For the other values of $T_r$, D was always 1.

This case is challenging because the average ionization state Z* changes by more than 10 ionization stages between "pure non-LTE" case, i.e., $T_r = 0$, or D=0, and the LTE case, i.e., $T_r = T_e$ and D=1. In order to obtain the full range charge states distribution, down to an ion fraction lower or equal to $10^{-6}$, we computed 19 charge states, from Xe$^{11+}$ (Xe XII in spectroscopic notation) to Xe$^{29+}$ (Xe XXX). The collisional radiative model was solved for the 20,375 relativistic configuration averages and the 2,070,392 fine structure levels belonging to the latter were computed, with configuration interaction within the complexes. Finally, the spectra were generated with the MUTA model.

We show that with 1% of the Planckian intensity, i.e., D = 0.01, at $T_r = T_e$ there is significant alteration of the charge distribution, with an average charge changing from Z* = 15.08 (D=0) to 16.69 (D=0.01). Further, the modification of the overall emission spectrum with $T_r$ is noticeable enough to provide an estimate of the latter if $T_e$ and $N_e$ can be measured separately, even for $T_r$ as small as 10 eV. We find that two different radiation fields yielding the same Z* of 17.89811, case A ($T_r = 40$ eV, D=1) and case B ($T_r = 100$ eV, D=0.032486), actually correspond to nearly, but not exactly, identical charge distributions. The ratio of intensity of two lines, belonging to the same Kr-like ion $Xe^{18+}$ differs sufficiently to be observable. We showed quantitatively that this is because there is a substantial difference in the photoexcitation of some levels due to the change in the radiation field. In addition, we showed that some levels are nearly unaffected by the radiation field, because their atomic structure does not allow a lower state from which they could be photo-excited .

In conclusion, we show that with a detailed atomic model, one can estimate the effect the total radiative field and also of its detailed shape. Conversely, the spectroscopic diagnostics of a plasma subject to a radiative shock can be in error if there is no detailed information on the radiation field.

**Acknowledgements**
We thank the Center for Radiative Shocks (CRASH) of the University of Michigan for partial support in form of the cooperation agreement Nº DE-FC52-08NA28616, and R. Paul Drake and Igor Sokolov for useful discussions.